\documentclass[12pt]{article}
\usepackage{qcircuit}
\usepackage{graphicx}
\usepackage[utf8]{inputenc}
\usepackage[dvips]{epsfig}
\usepackage{graphicx}
\usepackage[T1]{fontenc}
\usepackage{authblk}
\input{amssym}
\usepackage{float}
\usepackage{hyperref}
\date{}

\title{\bf Quantum simulation dynamics and circuit synthesis of FMO complex on an NMR quantum computer }
\author[1]{M. Mahdian\thanks{mahdian@tabrizu.ac.ir}}
\author[1]{H. Davoodi Yeganeh\thanks{h.yeganeh@tabrizu.ac.ir }}
\author[2]{A. Dehghani\thanks{adehghani@pnu.ac.ir}}

\affil[1]{Faculty of Physics, Theoretical and astrophysics
department , University of Tabriz, 51665-163 Tabriz, Iran}
\affil[2]{Department of Physics, Payame Noor University, P. O. Box
19395-3697, Tehran, Iran}
\begin{document}
\maketitle
\begin{abstract}
Recently, the dynamics simulation of light-harvesting complexes as an open quantum system, in the weak and strong coupling regimes, has received much attention. In this paper, we investigate a digital quantum simulation approach of the Fenna-Matthews-Olson (FMO) photosynthetic pigment-protein complex surrounded with a Markovian bath, i.e., memoryless, based on a nuclear magnetic resonance (NMR) quantum computer. For this purpose, we apply the decoupling(recoupling) method, which is turn off(on) the couplings and also Solovay-Kitaev techniques to decompose Hamiltonian and Lindbladians into efficient elementary gates on an NMR simulator. Finally, we design the quantum circuits for the unitary and non-unitary part due to the system-environment interactions of the open system dynamics.

\end{abstract}
\section{Introduction}

Real quantum systems interact as open systems with their surroundings, and understanding the dynamics of the interacting system, including many degrees of freedom in an environment, is one of the significant challenges in physics, chemistry, and biology.
Dynamic evolution of open systems treatment due to the decoherence, and dissipation effects isn't unitary and It's very complicated and often used to describe the dynamics of proximity like the Born-Markov approximation \cite{r15}.
Due to the exponential growth of variables, efficiently simulating quantum systems with complex many-body interactions are hard for classical computers.
For this reason, the idea of modeling a quantum computer to simulate large quantum systems was initially been suggested, by R. Feynman \cite{r1}, where he conjectured that the quantum computers might be able to carry the simulation more efficiently than the classical one.
Although, implementation of universal quantum computers for large sizes of open systems are not available yet. According to this, a quantum simulation was proposed to solve such an exponential explosion problem using a controllable quantum system \cite{r1}, and typically classified into two main categories, namely, analogs that are more scalable  \cite{l2} and digital that are more flexible and universal \cite{l1} . A lot of analytical and numerical methods had been employed to simulate the dynamics of open quantum systems \cite{r16, r17, r18, r19, r20, r21}. Different platforms have been
used for implementing quantum simulators, such as  ions trapped in the optical cavity \cite{r2, r3}, cold atoms in optical lattices
\cite{r4}, super-conducting qubits \cite{r5, r6}, photons \cite{r8}, quantum dots \cite{r9} and the spin qubits based on magnetic
resonance process \cite{r10, r11, r12, r13, r14}.
We will consider quantum simulation dynamics of photosynthetic light-harvesting
complexes as an open system in terms of some time sequence quantum gates. In all of the photosynthetic organisms,
light is absorbed  by pigments such as chlorophyll and carotene in
antenna complexes, and then this energy transfers as an electronic
excitation to a reaction center where charge separation occurs
through the processes. The FMO complex is made of three identical monomers where each monomer involves seven
bacteriochlorophyll molecules surrounded by a protein environment.
Neill Lambert et al. \cite{r22} introduced interesting progress in quantum
biology, where they performed an experimental and theoretical
studies on the photosynthesis such as the quantum coherent energy
transport, entanglement, and tests of quantumness. Decoherence in
biological systems are being studied in Ref. \cite{r23} and
principles of a noise-assisted transport as well as the origin of
long-lived coherences for the FMO complex in photosynthesis was
given. The exciton-energy transfer in light-harvesting complexes has
been investigated by various methods such as the Forster theory in a
weak molecular interaction limit or by the Redfield master equations
derived from Markov approximation in a weak coupling regime between
molecules and environment \cite{r24, r25, r26, r27, r28, r29}.
In general, the dynamics of an open quantum system can be divided into two categories. Markovian and non-Markovian dynamics that depend on the strength of the coupling between the system and the environment. In the weak coupling regime between the system and the environment, often proximity like the Born and Markov approximations are used. Non-Markovian effects on energy transfer dynamic are significant \cite{r30, r31}. Here, we ignore the non-Markovian
effects and the exact dynamics of the FMO complex can be described with the Markovian Lindblad master equation, which we will discuss in this article. It should be noticed that effective dynamics of the FMO complex is modeled by a Hamiltonian which describes the
coherent exchange of excitations among different sites, also local Lindblad terms that take into account the dissipation and dephasing
processes caused by a surrounding environment\cite{r38,N11}. In the one hand, simulations of the light-harvesting
complexes have considered, and a large number of various experimental and analytical studies have been done. For instance, numerical analysis of a spectral density based on the molecular dynamics has been studied in Refs. \cite{r31,r32,mahdian2016}, as well as the corresponding dynamics had been investigated based on a two-dimensional electronic spectroscopy \cite{r34}, superconducting qubits \cite{r35} and numerically and analytically simulations \cite{r36,mahdian2019}. On the other hand, nuclear spin systems are good candidates for a quantum simulator, because they include long coherence times and may be manipulated by complex sequences of radio frequency (RF) pulses, then they can be carried out easily using modern spectrometers. Because of the importance of the subject, we present an effective nuclear spin system using an NMR-based quantum simulator as a controllable quantum system that can be applied to simulate the Hamiltonian and Lindbladians of the FMO complex. We investigate a scheme based on the recoupling and decoupling methods
\cite{r37} ,which are particularly relevant to the connection of any
two nuclear spins using RF pulses in any selected time for
simulating  the Hamiltonian of FMO complex. Here, we assume the
Solovay-Kitaev decomposition strategy for single-qubit channels
\cite{r21} to simulate the non-unitary part of quantum master
equation and circuits obtained on the NMR quantum computation.

The paper is organized as follows. In Sec. 2, the FMO complex will be
introduced. In Sec. 3, simulation of Hamiltonian of the FMO complex
by an NMR simulator is expressed. In Sec. 4, corresponding
calculations to simulate the non-unitary part will be given.
Finally, Sec. 5 is devoted to some conclusions.
\section{FMO complex}
The FMO complex is generally constituted of multiple chromophores
which transform photons into exactions and transport to a reaction center.
The exciton dynamics for the light-harvesting system (e.g., in the FMO complex) is modelled by a Markovian master equation of
the form
\begin{equation}\label{density1 }
\dot{\rho}(t)=-i[H_{sys},\rho(t)]+\mathcal{L}_{deph}(\rho)+\mathcal{L}_{diss}(\rho),
\end{equation}
which contains the coherent exchange of excitation and local Lindblad terms \cite{r38,N11}.
The quantum coherent evolution is governed by a electronic tight-binding Hamiltonian with seven bacteriochlorophyll (BChl) sites in general form:
\begin{equation}\label{Hel }
    H_{sys} =\sum_{i=1}^{N=7} \varepsilon_{i}|i\rangle \langle i|+
    \sum_{i\neq j}^{N=7}\nu_{ij}(|i\rangle \langle j|+|j\rangle \langle
    i|),
\end{equation}
where $\varepsilon_{i}$ are the site energies and $\nu_{ij}$
is the Coulomb couplings of the transition densities of
the chromophores, often taken to be of the (Forster) dipole-dipole form.
We consider one excited molecule in time and others in ground states in every cycle so it goes as seven bacteriochorophyll sites space that $|i\rangle=|g_{1},...e_{i},...,g_{7}\rangle$ are the basis  denoting the excitations at site i for $i=1,...,7$.
To facilitate the comparison with NMR quantum computing, we rewrite the total Hamiltonian Eq.\ref{Hel } using Pauli matrices
\begin{equation}\label{hamior}
H_{sys}=\underbrace{\sum_{j=1}^7\epsilon_j\sigma^z_j }_{H_0}+
\underbrace{\sum_{j\neq l}^7 \nu_{jl} (\sigma^x_j\sigma^x_l +
\sigma^y_j\sigma^y_l)}_{H_I},
\end{equation}
where $H_0$ is single-qubit Hamiltonian and $H_I$ is long-term interaction.
For expressing the dynamics of non-unitary part, we assume
that the system affected by two distinct types of noise are called
the dissipative and dephasing processes. Dissipative effect passes
the excitation energy with rate $\Gamma_j$ to the environment and
dephasing process destroys the phase coherence with the rate $\gamma_j$
of the site $j^{th}$. Both of these processes can be describe using a
Markovian master equation with local dephasing and dissipation
terms. For the FMO complex in the Markovian master equation
approach, the dissipative and the dephasing processes are captured,
respectively, by the Lindblad super-operators as follows
\begin{equation}\label{diss}
\mathcal{L}_{diss}(\rho) = \sum_{j=1}^7\Gamma_j(-\sigma^+_j\sigma^-_j\rho - \rho\sigma^+_j\sigma^-_j + 2\sigma^-_j\rho\sigma^+_j),
\end{equation}
\begin{equation}\label{deph}
\mathcal{L}_{deph}(\rho) = \sum_{j=1}^7\gamma_j(-\sigma^+_j\sigma^-_j\rho - \rho\sigma^+_j\sigma^-_j + 2\sigma^+_j\sigma^-_j\rho\sigma^+_j\sigma^-_j).
\end{equation}

Finally, the total transfer of excitation is measured by the
population in the sink. In the next section, we introduce the recoupling
and decoupling method attached to simulate the Hamiltonian
of the FMO complex.
\section{Simulation of Hamiltonian with recoupling and decoupling method}
We use, the recoupling, and decoupling method with Hadamard
matrix's approach to simulate the Hamiltonian of FMO complex and
perform a specific coupling in the NMR quantum computation. The task of turning-off all the couplings is known as decoupling, and also
doing this for a selected subset of couplings is known as recoupling. These pulses are single-qubit operations that transfer Hamiltonian in time between two pulses so that unwanted couplings in a consecutive evolution cancel each other.
The decoupling part must be chosen as the Hadamard matrices $H(n)$, where we have used $\pm1$ instead of the positive and negative sign, respectively.
We use the $\sigma_x$ gates to control the signs of $\sigma_z$ for each spin over an equal interval, and we will achieve sign matrix $S_n$, where each column represents a time interval and each row indicates a qubit; for details, see Ref. \cite{r37}. When $H(n)$ does not necessarily exist, we start with $H(n)$ and finally take $S_n$. Similarly, for the recoupling part, we use a normalized Hadamard matrix $H(\bar{n})$, which has only $+$'s in the first row and column. Then, to implement selective recoupling between the $i^{th}$ and $j^{th}$ qubit, we exclude the first row and taking the second row of $H(\bar{n})$ to be the $i^{th}$ and $j^{th}$ qubit row of $ S_n$, also the other $n-2$ rows of  $ S_n$ can be chosen from the remaining rows of $H(\bar{n})$.

When a nuclear spin is placed in a static magnetic field$B_0$along the  z-direction, the dynamical
evolution will be dominated by the internal Hamiltonian $H=(1-\eta)\gamma B_0I_z=\frac{1}{2}\omega_0 \sigma_z$, where $\gamma$ is the nuclear gyromagnetic ratio,  $\eta$
is the chemical shift arising from the
partial shielding of $B_0$ by the electron cloud surrounding the nuclear spin, and $\omega_0=(1-\eta)\gamma B_0$ is the Larmor  frequency. Note, $I_z$  is the angular momentum
operator related to Pauli matrix $\sigma_z$.
For multiple-spin systems, heteronuclear spins are easily distinguished due to the
distinct  $\gamma$ and thus very different $\omega_0$ in the magnitude of hundreds of MHz, while
homonuclear spins are often individually addressed by the distinct $\eta$ due to different
local environments. Furthermore, the qubit-qubit interactions are the natural mediated
spin-spin interactions called Hamiltonian J-coupling terms. The
Hamiltonian for nearest-neighbor interaction is $H=\sum_l J_l \sigma_l^z\sigma_{l+1}^z$.
 Therefore, the total  Hamiltonian for a N-spin system is
\begin{equation}\label{isi1}
 H_{NMR} =\sum_{l=1}^N{1\over2}\omega_l\sigma^z_l + \sum_{l=1}^{N-1}J_l\sigma^z_l\sigma^z_{l+1},
\end{equation}
which forms a well-defined multi-qubit system used in most NMR quantum
computing \cite{N12}.

The Hamiltonian $H_{NMR}$ called Longitudinal Ising model in solid-state
systems and $J_l$ denotes a coupling strength between the $l^{th}$
and $(l+1)^{th}$ qubits. This Hamiltonian evolved in time by the
following unitary operator
\begin{equation}\label{isi2}
U_{NMR}\equiv U(\tau) = e^{-i{\tau\over4} H_{NMR}}.
\end{equation}
The key insight of this paper is that to use the NMR quantum simulator with $H_{NMR}$ to simulate the Hamiltonian of Eq.\ref{hamior} in two steps: the first step is to simulate single-qubit Hamiltonian $H_0$ and the second one is to simulate interaction part $H_I$. We then combine $H_0$ and $H_I$ to obtain the complete Hamiltonian of Eq.\ref{hamior}. The
detailed description of these steps is as given below.

\subsection{Simulation of $H_0$}

It is clear that the Hamiltonian of the FMO complex includes
seven-qubits, then the sign matrix should have seven rows. On the
other hand, the Hadamard matrix of order-7 does not exist; then we
consider a Hadamard matrix of order-8 to obtain a sign matrix $S_7$.
We obtain the time evolution for the first qubit and generalize it
to seven qubits, for this purpose by using the Eq.(\ref{isi2}), and
removing the last row of $H(8)$:
\begin{equation}
H(8) ={\pmatrix  {+1&+1&+1&+1&+1&+1&+1&+1 \cr +1&-1&+1&-1&+1&-1&+1&-1 \cr +1&+1&-1&-1&+1&+1&-1&-1 \cr +1&-1&-1&+1&+1&-1&-1&+1 \cr +1&+1&+1&+1&-1&-1&-1&-1 \cr +1&-1&+1&-1&-1&+1&-1&+1 \cr+1&+1&-1&-1&-1&-1&+1&+1 \cr +1&-1&-1&+1&-1&+1&+1&-1}},
\end{equation}
a possible sign matrix $S_7$ can be obtained as follows
\begin{equation}
S_7 ={\pmatrix{+&+&+&+&+&+&+&+ \cr +&-&+&-&+&-&+&- \cr +&+&-&-&+&+&-&- \cr +&-&-&+&+&-&-&+ \cr +&+&+&+&-&-&-&- \cr +&-&+&-&-&+&-&+ \cr+&+&-&-&-&-&+&+}}.
\end{equation}
In continuation with the following pulse sequence
$${U(\sigma^2_x\sigma^4_x\sigma^6_xU\sigma^2_x\sigma^4_x\sigma^6_x)(\sigma^3_x\sigma^4_x\sigma^7_
xU\sigma^3_x\sigma^4_x\sigma^7_x)(\sigma^2_x\sigma^3_x\sigma^6_x\sigma^7_xU\sigma^2_x\sigma^3_
x\sigma^6_x\sigma^7_x)}$$
$${(\sigma^5_x\sigma^6_x\sigma^7_xU\sigma^5_x\sigma^6_x\sigma^7_x)
(\sigma^2_x\sigma^3_x\sigma^5_xU\sigma^2_x\sigma^3_x\sigma^5_x)(\sigma^2_x\sigma^4_x\sigma^5_x\sigma^7_
xU\sigma^2_x\sigma^4_x\sigma^5_x\sigma^7_x)(\sigma^3_x\sigma^4_x\sigma^5_x\sigma^6_xU\sigma^3_x\sigma^4_
x\sigma^5_x\sigma^6_x)},$$ using the Eq.(\ref{isi2}) and the Paul
matrices, we have
\begin{equation}\label{isi9}
\left[e^{-i{\tau\over4}H_{NMR}}\sigma^2_x\sigma^3_x\sigma^4_x\sigma^5_x\sigma^6_x\sigma^7_xe^{-i{\tau\over4}H_{NMR}}\sigma^3_x\sigma^5_x\sigma^7_x\right]^2,
\end{equation}
which provides a time evolution of the first qubit, i.e.
\begin{equation}
u^z_1(\tau) = e^{-i{\tau\over2}\omega_1\sigma^z_1}.
\end{equation}
Quantum circuits to simulate $u^z_1(\tau)$ is shown in Figure 1.
Similarly, for seven qubits it can be written as
\begin{equation}
e^{-iH_0t} = \otimes^7_{l=1}u^z_l(\tau).
\end{equation}
with
\begin{equation}
u^z_l(\tau)= [e^{-i{\tau\over4}H_{NMR}}T_le^{-i{\tau\over4}H_{NMR}}\
{T}'_l]^2,
\end{equation}
where $\ {T}'_l =\ {\otimes}'_{j\neq l}\sigma^x_j$, $T_l =
\otimes_{j\neq l}\sigma^x_j$ with $l=1,2....6.$, and prime denotes
that if $j$ is odd (even) number, $l$ is considered as a even (odd)
number. Then, the time evolution of $H_0$ obtain at $\tau$= $4t$.
To test our quantum circuit, we use the Forest (pyQuil) software
platform. It is an open-source Python library developed by Rigetti
for constructing, analyzing, and running quantum programs
\cite{Smith}. We consider the initial state as $ |0000000\rangle$, and
then putting $\tau=1$, implement our circuit. The output of circuit
on Forest is
$$(0.8775825619-0.4794255386j)|00000000\rangle ,$$
which matches the theory, i.e. is equivalent to
$U_1^z(\tau)|0000000\rangle$. Since the time evolution occurred only
on the first qubit, it means that the recoupling has done( See
Appendix I for details and code).
\begin{figure}[h!]
\centering
\includegraphics[width=16cm]{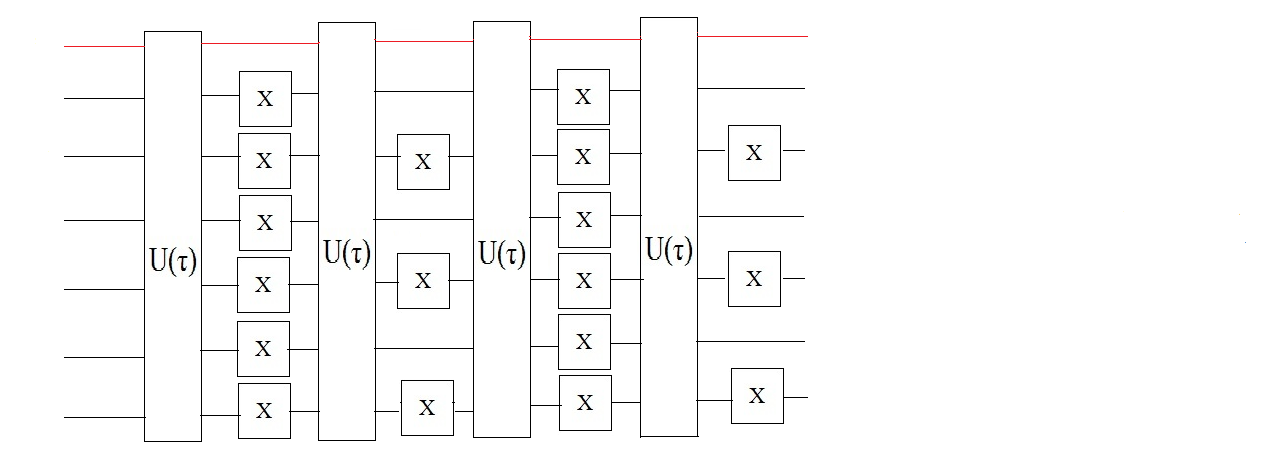}
\epsfysize=5cm
\epsfxsize=5cm
\caption{The quantum circuit to realize $U_1^z(\tau)$ from $U_{NMR}\equiv U(\tau)$. Red line shows the first qubit in the system and $X^i$ stand for gate $\sigma_{x}^i$}
\end{figure}

\subsection{ Simulation of $H_I$}
In this subsection, we simulate a time evolution of the Hamiltonian
$H_{I}$, by using a decoupling method. Similar to the case already
discussed in the previous section, the sign matrix $S_7$ is obtained by
a Hadamard matrix $H(8)$. We implement recoupling between the
$3^{rd}$ and $4^{th}$ qubits and finally expressed in the general
case for seven qubits. Firstly, we exclude the first row of $H(8)$
and take the second row of $H({8})$ to be the $3^{rd}$ and $4^{th}$
qubit row of $S_7$, Finally, five other rows of $S_7$ can be chosen
from the remaining rows of $H({8})$. A possible structure is
\begin{equation}
S_7 ={\pmatrix {+&+&-&-&+&+&-&- \cr +&-&-&+&+&-&-&+ \cr +&-&+&-&+&-&+&- \cr +&-&+&-&+&-&+&- \cr +&+&+&+&-&-&-&- \cr +&-&+&-&-&+&-&- \cr +&+&-&-&-&-&+&+}},
\end{equation}
in this case, the pulse sequence can be written as
$${U(\sigma^2_x\sigma^3_x\sigma^4_x\sigma^6_xU\sigma^2_x\sigma^3_x\sigma^4_x\sigma^6_x)(\sigma^1_x\sigma^2_x\sigma^7_xU\sigma^1_x\sigma^2_x\sigma^7_x)(\sigma^1_x\sigma^3_x\sigma^4_x\sigma^7_xU\sigma^1_x\sigma^3_x\sigma^4_x\sigma^7_x)}$$ $${(\sigma^5_x\sigma^6_x\sigma^7_xU\sigma^5_x\sigma^6_x\sigma^7_x)(\sigma^2_x\sigma^3_x\sigma^4_x\sigma^5_x\sigma^7_xU\sigma^2_x\sigma^3_x\sigma^4_x\sigma^5_x\sigma^7_x)(\sigma^1_x\sigma^2_x\sigma^5_x\sigma^6_xU\sigma^1_x\sigma^2_x\sigma^5_x\sigma^6_x)}$$ $${(\sigma^1_x\sigma^3_x\sigma^4_x\sigma^5_x\sigma^6_xU\sigma^1_x\sigma^3_x\sigma^4_x\sigma^5_x\sigma^6_x)}.$$
Along with the identity $\sigma_x^2=I$, it can be recast into:
\begin{equation}
e^{-i\tau j_3 \sigma^z_3 \sigma^z_4}= U^{zz}_{3\: 4}(\tau)=
U(\sigma^1_x\sigma^2_x\sigma^4_x\sigma^5_x\sigma^6_x\sigma^7_xU\sigma^1_x\sigma^5_x\sigma^7_x)U(\sigma^1_x\sigma^2_x\sigma^4_x\sigma^5_x\sigma^6_x\sigma^7_xU\sigma^1_x\sigma^5_x\sigma^7_x).
\end{equation}
Note we consider $x-x$ and $y-y$ interaction for two
nearest-neighbor-interacting qubits, and then by using the
single-qubit operations, we obtain:
\begin{equation}\label{isi11}
U^{ xx +yy}_{3\: 4}(\tau) = e^{-i\tau H^{xy}_{3\: 4}} =
e^{i{\pi\over 4} \sigma^y_3 \sigma^y_4}U^{zz}_{3 4}e^{-i{\pi\over 4}
\sigma^y_3 \sigma^y_4}e^{i{\pi\over 4} \sigma^x_3
\sigma^x_4}U^{zz}_{3\: 4}e^{-i{\pi\over 4} \sigma^x_3 \sigma^x_4},
\end{equation}
where we have used the notation $ H^{xy}_{3\: 4} = J_3(\sigma^x_3
\sigma^x_4 + \sigma^y_3 \sigma^y_4)$. Also, for seven qubits it can
be written, generally, as follows\\\\ $U^{ xx +yy}_{l\: l+1}(\tau)
=e^{-i\tau H^{xy}_{l\: l+1}}= e^{i{\pi\over 4} \sigma^y_l
\sigma^y_{l+1}}U^{zz}_{l\: l+1 }e^{-i{\pi\over 4} \sigma^y_l
\sigma^y_{l+1}}e^{i{\pi\over 4} \sigma^x_l
\sigma^x_{l+1}}U^{zz}_{l\: l+1}e^{-i{\pi\over 4} \sigma^x_l
\sigma^x_{l+1}},$\\\\ where $\tau = t$, $ J_l= 2\nu_{jl} $ with $ l=
1,2...6$ and considering $ H^{xy}_l= J_l(\sigma^x_l\sigma^x_{l+1} +
\sigma^y_l\sigma^y_{l+1})$. Quantum circuits to simulate $U^{ xx
+yy}_{3\: 4}(\tau)$ is shown in figure 2. Similar to the previous
one, by implementing circuit on the Forest's software
platform  by:\\
$$(0.0690086667-0.9957193521j)|00000000\rangle$$
$$+(-0.0346499137+0.0177942584j)|00000100\rangle$$
$$+(0.0273581824-0.0386110549j)|00001000\rangle$$
$$+(-0.0034121934+0.0035492127j)|00001100\rangle,$$ it is
approximately equal with direct calculation of $ U^{ xx
+yy}_{3\:4}(\tau)|0000000\rangle$. It is easily seen that coupling
between the $3^{th}$ and $4^{th}$ site is conserved, i.e. the
decoupling yield with high efficiency. With all nearest-neighbor
coupling operators $U^{zz}_{l\: l+1}$ and $U^{ xx +yy}_{l\: l+1}$
being simulated, one can extend them to the long-range interactions
in a straightforward manner. Since, both the Hamiltonian $H_0$
and$H_I$ are available, then the total Hamiltonian $H$ can be
obtained by the Trotters formula \cite{r39}:
\begin{equation}
e^{-iHt} = e^{-iH_0t}e^{-iH_It} + o(t^2).
\end{equation}
Here we have used $(H_{NMR})$ as a Hamiltonian simulator,
which can be used instead of other Hamiltonian
such as the transverse Ising model, the $XY$ and
Heisenberg model.
\begin{figure}[h!]
\centering
\includegraphics[width=16cm]{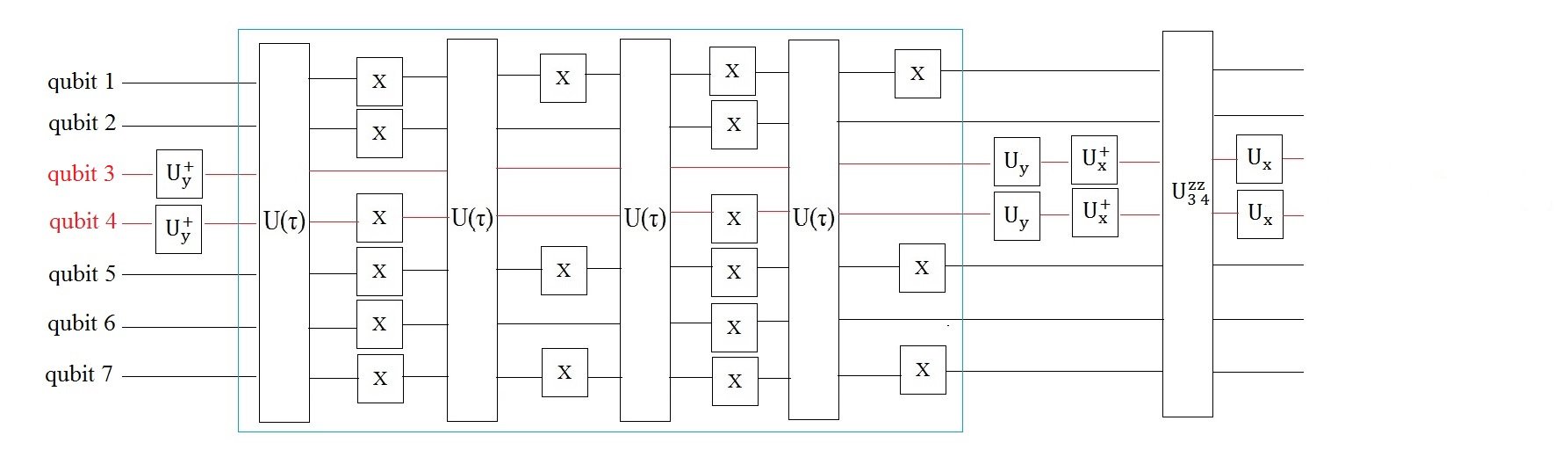}
\caption{The quantum  circuit is characterized the XY exchange
interaction on the qubits  third and fourth,  $U^{ xx +yy}_{3\: 4}(\tau)$ .
Rectangular boxes connecting to  simulating  $ U^{zz}_{3\: 4}(\tau) $  directly from $ U_{NMR}\equiv U(\tau)$}
\end{figure}

\section{ Quantum simulation of the non-unitary part}
It can be deduced from the non-unitary part of the FMO
complex that each monomer has seven bacteria and considered as a
system of seven qubits. We consider one of them which interacts with their surroundings and study its dynamics with the single-qubit
channels approach. So we restrict ourselves to single-qubit states
and begin by recalling some geometric properties of them \cite{r40}.
In general, every density matrix $\rho$ can be written in terms of
the standard bases, $\{I, \sigma^x , \sigma^y , \sigma^z\}$, as
$\rho = \frac{1}{2}(I + \bf r. \sigma)$ with ${\bf r}\in \Bbb{R}^3 $
and $|\bf r|$=$1$. Each single qubit quantum channel can then be
represented in this basis by a unique $4 \times 4$ matrix $ T=
{\pmatrix{1 & \bf 0 \cr \bf m & M}}$, where $M$ is a $3 \times 3$
matrix and  $\bf 0$, $\bf m$ denote row and column vectors
respectively. The density matrix through the action of a channel
will change as follows.
\begin{equation}
T(\rho) = \ {\rho}' =\frac{1}{2}(I + {\ {r}'} . \bf {\sigma)},
\end{equation}
with $ {\ {r}'} = {M.r +\bf{m} }$ and the channel, $T$, is
considered as an affine map, i.e.
\begin{equation}
T = {\pmatrix {1&0&0&0 \cr 0&\Lambda_1&0&0 \cr 0&0&\Lambda_2&0 \cr m_3&0&0&\Lambda_3 }},
\end{equation}
by
\begin{equation}
M = {\pmatrix{\Lambda_1&0&0 \cr 0&\Lambda_2&0 \cr 0&0&\Lambda_3 }}.
\end{equation}
Also, the matrix $T$ can be rewritten in the following form
\begin{equation}
T = {\pmatrix {1&0&0&0 \cr 0&\cos{\upsilon}&0&0 \cr 0&0&\cos{\mu}&0 \cr \sin{\upsilon}\sin{\mu}&0&0&\cos{\upsilon}\cos{\mu} }},
\end{equation}
in this case, the Kraus operators will be given by
\begin{equation}
K_1 = {\pmatrix {\cos{\beta}&0 \cr 0&\cos{\alpha} }},
\end{equation}
\begin{equation}
K_2 = {\pmatrix {0&\sin{\alpha} \cr \sin{\beta}&0 }},
\end{equation}
where $\alpha = \frac{1}{2}(\mu + \upsilon)$ and $\beta = \frac{1}{2}(\mu - \upsilon)$. For one qubit state the Eq. (\ref{diss}) becomes
\begin{equation}
\mathcal{L}_{diss}(\rho) = \Gamma(-\sigma^+\sigma^-\rho - \rho\sigma^+\sigma^- + 2\sigma^-\rho\sigma^+),
\end{equation}
by using the damping basis methods \cite{r41,r42} (details are given in the Appendix) and considering
$$\Lambda_{1,2} =e^{-4\Gamma t}, \Lambda_3 =e^{-8\Gamma t}, m_3 = e^{-8\Gamma t} - 1,$$
one can find ${\rho}'$ as the following equation
\begin{equation}\label{den1}
\ {\rho}' = \frac{I + (-1 +e^{-8\Gamma t}(1 + r_z))\sigma_z +
r_x\sigma_x e^{-4\Gamma t} +  r_y\sigma_y e^{-4\Gamma t}}{2}.
\end{equation}
Any single-qubit channel $T$ (CPTP map) can be simulated with one ancillary qubit, one CNOT, and four single-qubit
operations\cite{r21}. The two rotation operations are applied to
cover the Kraus operator's action, and another single-qubit operation
is used only to diagonalize the matrix $M$. We want to design a
circuit to simulate the non-unitary part dynamics of the FMO complex on the NMR computer. So recalling some properties of the NMR quantum computation \cite{r43}, we shall consider a physical system which consists of a solution of identical molecules. Each molecule has N magnetically inequivalent nuclear spins, which serve as qubits.
Nuclear spins interact via a dipole-dipole coupling or indirect
coupling mediated by electrons. In any case, in the presence of a
strong external magnetic field, only the secular parts are important
\cite{r37}. Single qubit operations can be induced by the RF
magnetic fields, oriented in the $x-y$ plane perpendicular to the
static field. The RF pulse can be selectively addressed spin $i$ by
an oscillator at angular frequency $\omega_i$. The general form of
single-qubit gates in quantum information processing may become as
the RF pulse along the $\hat{n}$-axis induces the rotation operator
$e^{-i{t_{pw}\over2} \sigma.\hat{n}}$ where $t_{pw}$ is proportional
to the pulse duration (pulse width) and amplitude. For example $X =i
e^{-i {\pi \over2}\sigma_x}$ can be considered as a single $\pi-$
pulse around $\hat{x}$-axes, then the Hadamard gate $H=
ie^{-i{\pi\over2}\sigma_x}e^{-i{\pi \over2}\sigma_y}$ can create a
${\pi\over2}$-pulse around $ \hat{y}$-axes followed by a $\pi$-pulse
around $\hat{x}$-axes, too. Coupled logic gates can be naturally
performed by a time evolution of the system. It can be assumed that
the individual coupling term can be selectively turned on to perform
a coupled operation between $i^{th}$ and $j^{th}$ qubits, next
turning on the coupling term $g_{ij}\sigma^i_z\otimes \sigma^j_z$
for time $t$, leads to the evolution of logic gate
$e^{-itg_{ij}\sigma^i_z\otimes \sigma^j_z}$. Together with setting
all of single-qubit transformations, the $CNOT_{ij} = (I_i \otimes
H_j)Cz(I_i \otimes H_j)$ fulfills a requirement for universality.
Returning to the original problem and starting by the following
assumptions:
$$ \cos\alpha = e^{-4\Gamma t},\quad \cos\beta =1, \quad \sin\alpha = \sqrt{1- e^{-8\Gamma t}}, \quad sin\beta=0,$$
the kraus operators are obtained as follows
\begin{equation}
K^{diss}_1 = {\pmatrix{1&0 \cr 0& e^{-4\Gamma t}}},
\end{equation}
\begin{equation}
K^{diss}_2 = {\pmatrix{0& \sqrt{1- e^{-8\Gamma t}} \cr 0&0}}.
\end{equation}
As mentioned above an action of the Kraus operators can be
represented through the rotations $ R_y(2\delta_1(2))$ = $ e^{-
i\sigma_y \gamma_1(2)}$ with $2\delta_1 = \beta - \alpha + {\pi
\over 2}$ and $2\delta_2 = \beta + \alpha - {\pi \over 2} $ in a
quantum circuit. For implementing the rotations $ R_y(2\delta_1)$
and $ R_y(2\delta_2)$, respectively, we set
\begin{equation}
t_{pw} =2\delta_1 = -Arc\cos(-4\Gamma t) + {\pi \over 2}\qquad and
\qquad \hat{n}= \hat{y},\end{equation} and
\begin{equation}
t_{pw} =2\delta_2 = Arc\cos(-4\Gamma t) - {\pi \over 2} \qquad and
\qquad  \hat{n}=\hat{y}.\end{equation} Along with the CNOT gate and
above mentioned preliminaries, we can obtain a quantum circuit for
implementation of the quantum channel of T for dissipation process that
shown in Figure. 3. For the dephasing process, a straightforward
calculations of the Kraus operators leads to
\begin{equation}
K^{deph}_1 = {\pmatrix{{-1\over2}e^{-2\gamma t}&0 \cr 0& {1\over2}e^{-2\gamma t}}},
\end{equation}
\begin{equation}
K^{deph}_2= {\pmatrix{0&  \sqrt{1-{1\over2} e^{-2\gamma t}} \cr \sqrt{1-{-1\over2}e^{-2\gamma t}}&0}},
\end{equation}
similarly, this process is also implemented based on the NMR
simulator. In comparing with the implementation process of the
$R_y(2\gamma_1)$ and $R_y(2\gamma)$ we choose here $t_{pw}=
2\gamma_1= [Arc \cos\left(-{e^{-2\gamma t}\over2}\right)]/2$ and
$\hat{n} =\hat{y}$ and $t_{pw} = 2\gamma_2=-{\pi\over2}$ and
$\hat{n}=\hat{y}$ respectively.
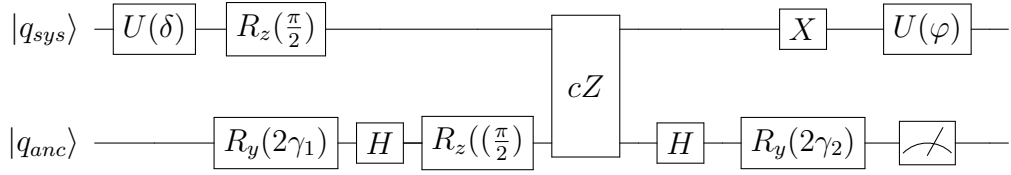
\begin{figure} [t]
\centerline{
\Qcircuit @C=0.6em @R=1.95em {\lstick{|q_{sys}\rangle}
& \gate{U(\delta)} & \gate{R_z({\pi \over2})} &    \qw & \qw &  \multigate{1}{cZ} & \qw  & \qw   &  \qw   & \gate{X} & \gate{U(\varphi)} & \qw & \qw \\
\lstick{|q_{anc}\rangle} &   \qw & \gate{R_y(2\gamma_1)} & \gate{H}    & \gate{R_z(({\pi \over2})} \qw & \ghost{cZ} & \qw & \gate{H} & \qw & \gate{R_y(2\gamma_2)} & \meter & \qw & \qw &}}
\caption{The quantum  circuit to implement the simulation of one qubit dynamics on \textbf{the} nuclear spin system. The unitary operators $U(\delta)$ and $U(\phi)$ serve diagonalize the channel and $|q_{sys}\rangle$ , $|q_{anc}\rangle$ denote \textbf{the} state of system and ancilla qubit respectively.}
\end{figure}
\section{Conclusion}
In this paper, we investigated a quantum simulation of the
FMO complex dynamics by nuclear spin systems. We employed recoupling
and decoupling methods to simulate the Hamiltonian of FMO complex,
then for quantum simulation of the non-unitary part dynamics of FMO
complex with the single-qubit channel's a quantum circuit had obtained.
Finally, the obtained circuit implements an NMR quantum computation
based on Forest's software platform (pyQuil). The output of
the pyQuil code is compatible with direct calculation, too. Also,
dynamical simulation of the FMO complex optimized with currently
available technology. However, we will study the non-Markovian dynamics of the FMO complex base on NMR quantum computation in the future.\\
\begin{appendix}
{\bf Appendix I}\\
As mentioned above pyQuil is an open-source Python library developed by Rigetti for constructing, analyzing, and running quantum programs. It is built on top of Quil, an open quantum instruction language (or simply quantum language), designed specifically for near-term quantum computers and based on a shared classical/quantum memory model \cite{Smith}. By using this instruction we implement the Eq.(\ref{isi9}) on pyQuil  by below code:
\\

from pyquil.quil import\\
from pyquil.api import WavefunctionSimulator\\
for i in list1:\\
    p+=H(i),CPHASE(np.pi,i+1,i),RX(0.25,i),CPHASE(np.pi,i+1,i),H(i),RZ(0.25,i)\\
for i in list1:\\
    if i>0:\\
        p+=X(i)\\
p+=X(2),X(4),X(6)\\
for i in list1:\\
    p+=H(i),CPHASE(np.pi,i+1,i),RX(0.25,i),CPHASE(np.pi,i+1,i),H(i),RZ(0.25,i)\\
for i in list1:\\
    if i>0:\\
        p+=X(i)\\
for i in list1:\\
    p+=H(i),CPHASE(np.pi,i+1,i),RX(0.25,i),CPHASE(np.pi,i+1,i),H(i),RZ(0.25,i)\\
p+=X(2),X(4),X(6)
\#
\#
\#
\#

Similarly, the code  to implement of  Ee.(\ref{isi11}) can be written straightforward.

{\bf Appendix II}\\
To solve the master equation which $\mathcal{L}$ is the generator of a semigroup of a quantum channel at first, we must find left and right eigen-operators $L_k$ and $R_k$ which satisfying the following condition:
\begin{equation}
L_k\mathcal{L} = \lambda_{(k,j)} L_k, \qquad
\end{equation}

\begin{equation}
R_k\mathcal{L} = \lambda_{(k,j)}R_k,
\end{equation}
\noindent
By using  the left action of superopertor that defined as $ \emph{Tr}[(\ell(\rho))O] =\emph{Tr}[(O\ell)\rho]$ for arbitrary Hermitian operator $O$ and any density matrix can find  that $\emph{Tr}[L_k R_m]$ = $\delta_{k m}$ and $\lambda_{(L,k)} =\lambda_{(R,k)}$ where $\emph{Tr}$ refers to the usual trace, so that initial state writing in damping base method such \cite{r41, r42}
\begin{equation}
\rho(0) = \sum_{k} \emph{\emph{Tr}}[L_k\rho(0)]R_k,
\end{equation}
and
\begin{equation}\label{den2}
\rho(t) = e^{\mathcal{L} t}[\rho(0)] = \sum_{k}\emph{Tr}[L_k\rho(0)]\Lambda_k R_k,
\end{equation}
where $\Lambda_k = e^{\lambda_k t}$. So for solving equation (23) we utilize these set \{$I$, $\sigma_z$ , $\sigma^+$ and $\sigma^-$\} as right eigen-operators, we obtain:
\begin{equation}
\mathcal{L}_R(I) = -8\Gamma \sigma_z,
\end{equation}
\begin{equation}
\mathcal{L}_R(\sigma_z) = -8\Gamma \sigma_z,
\end{equation}
\begin{equation}
\mathcal{L}_R(\sigma^+) = -4\Gamma \sigma^+,
\end{equation}
\begin{equation}
\mathcal{L}_R(\sigma^-) = -4\Gamma \sigma^-,
\end{equation}
For left  eigen-operators action we consider an appropriate set of operators \{ $(I - \sigma_z)$, $\sigma_z$, $\sigma^+$ and $\sigma^-$\}
\begin{equation}
\mathcal{L}_L(I - \sigma_z) = 0,
\end{equation}
\begin{equation}
\mathcal{L}_L(\sigma_z) = -8\Gamma \sigma_z,
\end{equation}
\begin{equation}
\mathcal{L}_L(\sigma^+) = -4\Gamma \sigma^+,
\end{equation}
\begin{equation}
\mathcal{L}_L(\sigma^-) = -4\Gamma \sigma^-,
\end{equation}
so $    \emph{Tr}[L_1\rho(0)] =1/2(1-r_z)$ ,$\emph{Tr}[L_2\rho(0)] = 1/2r_z$ , $\emph{Tr}[L_3\rho(0)] = 1/4(r_x  + i r_y)$  and $Tr\emph{}[L_4\rho(0)] =  1/4(r_x  - i r_y)$
Using Eq.(\ref{den2}) we can easily obtain Eq.(\ref{den1}).\\
\\
\\

\end{appendix}
\bibliographystyle{ieeetr}
\bibliography{references}
\end{document}